# Derivation of a novel efficient supervised learning algorithm from cortical-subcortical loops

**Ashok Chandrashekar[1]* and Richard Granger[2]**

[1] Department of Computer Science, Dartmouth College, Hanover, NH, USA
[2] Psychological and Brain Sciences, Thayer School of Engineering and Computer Science, Dartmouth College, Hanover, NH, USA

***Correspondence:**
Ashok Chandrashekar, Department of Computer Science, Dartmouth College, Hinman Box 6211, Hanover, NH 03775, USA.
e-mail: ashok.chandrashekar@dartmouth.edu

Although brain circuits presumably carry out powerful perceptual algorithms, few instances of derived biological methods have been found to compete favorably against algorithms that have been engineered for specific applications. We forward a novel analysis of a subset of functions of cortical–subcortical loops, which constitute more than 80% of the human brain, thus likely underlying a broad range of cognitive functions. We describe a family of operations performed by the derived method, including a non-standard method for supervised classification, which may underlie some forms of cortically dependent associative learning. The novel supervised classifier is compared against widely used algorithms for classification, including support vector machines (SVM) and k-nearest neighbor methods, achieving corresponding classification rates – at a fraction of the time and space costs. This represents an instance of a biologically derived algorithm comparing favorably against widely used machine learning methods on well-studied tasks.

## 1. INTRODUCTION

Distinct brain circuit designs exhibit different functions in human (and other animal) brains. Particularly notable are studies of the basal ganglia (striatal complex), which have arrived at closely-related hypotheses, from independent laboratories, that the system carries out a form of reinforcement learning (Sutton and Barto, 1990; Schultz et al., 1997; Schultz, 2002; Daw, 2003; O'Doherty et al., 2003; Daw and Doya, 2006); despite ongoing differences in the particulars of these approaches, their overall findings are surprisingly concordant, corresponding to a still-rare instance of convergent hypotheses of the computations produced by a particular brain circuit. Models of thalamocortical circuitry have not yet converged to functional hypotheses that are as widely agreed-on, but several different approaches nonetheless hypothesize the ability of thalamocortical circuits to perform unsupervised learning, discovering structure in data (Lee and Mumford, 2003; Rodriguez et al., 2004; Granger, 2006; George and Hawkins, 2009). Yet thalamocortical and striatal systems do not typically act in isolation; they are tightly connected in cortico-striatal loops such that virtually each cortical area interacts with corresponding striatal regions (Kemp and Powell, 1971; Alexander and DeLong, 1985; McGeorge and Faull, 1988). The resulting cortico-striatal loops constitute more than 80% of human brain circuitry (Stephan et al., 1970, 1981; Stephan, 1972), suggesting that their operation provides the underpinnings of a very broad range of cognitive functions.

We forward a new hypothesis of the interaction between cortical and striatal circuits, carrying out a hybrid of unsupervised hierarchical learning and reinforcement, together achieving a cortico-striatal loop algorithm that performs a number of distinct operations of computational utility, including supervised and unsupervised classification, search, object and feature localization, and hierarchical memory organization. For purposes of the present paper we focus predominantly on the particular task of supervised learning.

Traditional supervised learning methods typically identify class boundaries by focusing primarily on the class labels, whereas unsupervised methods discover similarity structure occurring within a dataset; two distinct tasks with separate goals, typically carried out by distinct algorithmic approaches.

Widely used supervised classifiers such as support vector machines (Vapnik, 1995), supervised neural networks (Bishop, 1996), and decision trees (Breiman et al., 1984; Buntine, 1992), are so-called discriminative models, which learn separators between categories of sample data without learning the data itself, and without illuminating the similarity structure within the data set being classified.

The cortico-striatal loop (CSL) algorithm presented here is "generative," i.e., it is in the category of algorithms that models data occurring within each presented class, rather than seeking solely to identify differences between the classes (as would a "discriminative" method). Generative models are often taken as performing excessive work in cases where the only point is to distinguish among labeled classes (Ng and Jordan, 2002). The CSL method may thus be taken as carrying out more tasks than classification, which we indeed will see it does. Nonetheless, we observe the behavior of the algorithm in the task of classification, and compare it against discriminative classifiers such as support vectors, and find that even in this restricted (though very widely used) domain of application, the CSL method achieves comparable classification as discriminative models, and uses far less computational cost to do so, despite carrying out the additional work entailed in generative learning.

The approach combines the two distinct tasks of unsupervised classification and reinforcement, producing a novel method for yet another task: that of supervised learning. The new method identifies supervised class boundaries, as a byproduct of uncovering structure in the input space that is independent of the supervised labels. It performs solely unsupervised splits of the data into similarity-based clusters. The constituents of each subcluster are checked to see whether or not they all belong to the same intended supervised category. If not, the algorithm makes another unsupervised split of the cluster into subclusters, iteratively deepening the class tree. The process repeats until all clusters contain only (or largely) members of a single supervised class. The result is the construction of a hierarchy of mostly mixed classes, with the leaves of the tree being "pure" categories, i.e., those whose members contain only (or mostly) a single shared supervised class label.

Some key characteristics of the method are worth noting.

- Only unsupervised splits are performed, so clusters always contain only members that are similar to each other.
- In the case of similar-looking data that belong to distinct supervised categories (e.g., similar-looking terrains, one leading to danger and one to safety), these data will constitute a difficult discrimination; i.e., they will reside near the boundary that partitions the space into supervised classes.
- In cases of similar data with different class labels, i.e., difficult discriminations, the method will likely perform a succession of unsupervised splits before happening on one that splits the dangerous terrains into a separate category from the safe ones.

In other words, the method will expend more effort in cases of difficult discriminations. (This characteristic is reminiscent of the mechanism of support vectors, which identify those vectors near the intended partition boundary, attempting to place the boundary so as to maximize the distance from those vectors to the boundary.) Moreover, in contrast to supervised methods that provide expensive, detailed error feedback at each training step (instructing the method as to which supervised category the input should have been placed in), the present method uses feedback that is comparatively far more inexpensive, consisting of a single bit at each training step, telling the method whether or not an unsupervised cluster is yet "pure"; if so, the method stops for that node; if not, the method performs further unsupervised splits.

This deceptively simple mechanism not only produces a supervised classifier, but also uncovers the similarity structure embedded in the dataset, which competing supervised methods do not. Despite the fact that competing algorithms (such as SVM and Knn) were designed expressly to obtain maximum accuracy at supervised classification, we present findings indicating that even on this task, the CSL algorithm achieves comparable accuracy, while requiring significantly less computational resource cost.

In sum, the CSL algorithm, derived from the interaction of cortico-striatal loops, performs an unorthodox method that rivals the best standard methods in classification efficacy, yet does so in a fraction of the time and space required by competing methods.

## 2. CORTICO-STRIATAL LOOPS

The basal ganglia (striatal complex), present in reptiles as well as in mammals, is thought to carry out some form of reinforcement learning, a hypothesis shared across a number of laboratories (Sutton and Barto, 1990; Schultz et al., 1997; Schultz, 2002; Daw, 2003; O'Doherty et al., 2003; Daw and Doya, 2006). The actual neural mechanisms proposed involve action selection through a maximization of the corresponding reward estimate for the action on the task (see Brown et al., 1999; Gurney et al., 2001; Daw and Doya, 2006; Leblois et al., 2006; Houk et al., 2007 for a range of views on action selection). This reward estimation occurs in most models of the striatum through the regulation of the output of the neurotransmitter dopamine. Therefore, in computational terms we can characterize the functionality of the striatum as an abstract search through the space of possible actions, guided by dopaminergic feedback.

The neocortex and thalamocortical loops are thought to hierarchically organize complex fact and event information, a hypothesis shared by multiple researchers (Lee and Mumford, 2003; Rodriguez et al., 2004; Granger, 2006; George and Hawkins, 2009). For instance, in Rodriguez et al. (2004) the anatomically recognized "core" and "matrix" subcircuits are hypothesized to carry out forms of unsupervised hierarchical categorization of static and time-varying signals; and in Lee and Mumford (2003), George and Hawkins (2009), Riesenhuber and Poggio (1999), and Ullman (2006) and many others, hypotheses are forwarded of how cortical circuits may construct computational hierarchies; these studies from different labs propose related hypotheses of thalamocortical circuits performing hierarchical categorization.

It is widely accepted that these two primary telencephalic structures, cortex and striatum, do not act in isolation in the brain; they work in tight coordination with each other (Kemp and Powell, 1971; Alexander and DeLong, 1985; McGeorge and Faull, 1988). The ubiquity of this repeated architecture (Stephan et al., 1970, 1981; Stephan, 1972) suggests that cortico-striatal circuitry underlies a very broad range of cognitive functions. In particular, it is of interest to determine how semantic cortical information could provide top-down constraints on otherwise too-broad search during (striatal) reinforcement learning (Granger, 2011). In the present paper we study this interaction in terms of subsets of the leading extant computational hypotheses of the two components: thalamocortical circuits for unsupervised learning and the basal ganglia/striatal complex for reinforcement of matches and mismatches. If these bottom-up analyses of cortical and striatal function are taken seriously, it is of interest to study what mechanisms may emerge from the interaction of the two mechanisms when engaged in (anatomically prevalent) cortico-striatal loops.. We adopt straightforward and tractable simplifications of these models, to study the operations that arise when the two are interacting. **Figure 1** illustrates a hypothesis of the functional interaction between unsupervised hierarchical clustering (uhc; cortex) and match-mismatch reinforcement (mm; striatal complex), constituting the integrated mechanism proposed here.

The interactions in the simplified algorithm are modeled in part on mechanisms outlined in Granger (2006): a simplified model of thalamocortical circuits produces unsupervised clusters of the input data; then, in the CSL model, the result of the clustering, along with the corresponding supervised labels, are examined by a simplified model of the striatal complex. The full computational models of the thalamocortical hierarchical clustering and sequencing circuit and striatal reinforcement-learning circuit yield interactions that are under ongoing study, and will, it is hoped, lead to further derivation of additional algorithms. For the present paper, we use just small subsets of the hypothesized functions of these structures: solely the hypothesized hierarchical clustering function of the thalamocortical circuit, and a very-reduced subset of the reinforcement-learning capabilities of the striatal complex, such that it does nothing more than compare (match / mismatch) the contents of a proposed category, and return a single bit corresponding to whether the contents all have been labeled as "matching" each other (1) or not (0). This very-reduced RL mechanism can be thought of simply as rewarding or punishing a category based on its constituents. In particular the proposed simplified striatal mechanism returns a single bit (correct/incorrect) denoting whether the members of a given unsupervised cluster all correspond to the same supervised "label." If not, the system returns a "no" ("–") to the unsupervised clustering mechanism, which in turn iterates over the cluster producing another, still unsupervised, set of subclusters of the "impure" cluster. The process continues until each unsupervised subcluster contains members only (or mostly, in a variant of the algorithm) of a single category label.

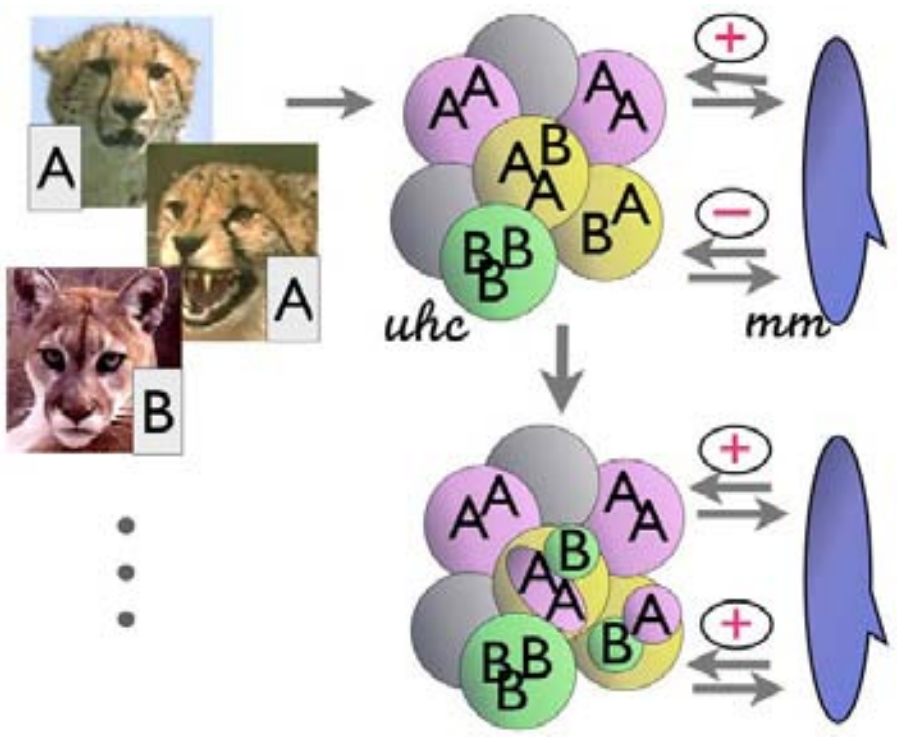

**FIGURE 1 | Simplified hypothesis of the computations of cortico-striatal interaction.** (Left) Inputs (e.g., images) arrive together with associated labels (A, B, etc.). (Top right) Unsupervised hierarchical clustering (*uhc*) categorizes the inputs by their similarity, without initial regard to labels. (Far right) The match–mismatch mechanism (*mm*) matches the class membership labels within each of these unsupervised categories. If the members of a category all have the same (matching) labels, a "+" is returned; if there are mismatches, a "−" is returned. (Bottom right) In the latter case, the clustering mechanism iteratively deepens the hierarchical tree by creating subclusters (e.g., central node in the diagram); these in turn are checked for label consistency, as before. The process iterates until the leaf nodes of the unsupervised tree contain only category members of a single label. (See text).

In sum, the mechanism uses only unsupervised categorization operations, together with category membership tests. These two mechanisms result in the eventual iterative arrival at categories whose members can be considered in terms of supervised classes.

Since only unsupervised splits are performed, categories (clusters) always contain only members that are similar to each other. The tree may generate multiple terminal leaves corresponding to a given class label; in such cases, the distinct leaves correspond to dissimilar class subcategories, eventually partitioned into distinct leaf nodes. The mechanism can halt rapidly if all supervised classes correspond to similarity-based clusters; i.e., if class labels are readily predictable from their appearance. This corresponds to an "easy" discrimination task. When this is not the case, i.e., in instances where similar-looking data belong to different labeled categories (e.g., similar mushrooms, some edible and some poisonous), the mechanism will be triggered to successively subdivide clusters into subclusters, as though searching for the characteristics that effectively separate the members of different labels.

In other words, less work is done for "easy" discriminations; and only when there are difficult discriminations will the mechanism perform additional steps. The tree becomes intrinsically unbalanced as a function of the lumpiness of the data: branches of the tree are only deepened in regions of the space where the discriminations are difficult, i.e., where members of two or more distinct supervised categories are close to each other in the input space. This property is reminiscent of support vectors, which identify boundaries in the region where two categories are closest (and thus where the most difficult discriminations occur).

A final salient feature of the mechanism is its cost. In contrast to supervised methods, which provide detailed, expensive, error feedback at each training step (telling the system not only when a misclassification has been made but also exactly which class should have occurred), the present method uses feedback that by comparison is extremely inexpensive, consisting of a single bit, corresponding to either "pure" or "impure" clusters. For pure clusters, the method halts; for impure clusters, the mechanism proceeds to deepen the hierarchical tree.

As mentioned, the method is generative, and arrives at rich models of the learned input data. It also produces multiclass partitioning as a natural consequence of its operation, unlike discriminative supervised methods which are inherently binary, requiring extra mechanisms to operate on multiple classes.

Overall, this deceptively simple mechanism not only produces a supervised classifier, but also uncovers the similarity structure embedded in the dataset, which competing supervised methods do not. The terminal leaves of the tree provide final class information, whereas the internal nodes provide further information: they are mixed categories corresponding to meta labels (e.g., superordinate categories; these also can provide information about which classes are likely to become confused with one another during testing.

In the next section we provide an algorithm that retains functional equivalence with the biological model for supervised learning described above while abstracting out the implementation details of the thalamocortical and striatal circuitry. Simplifying the implementation enables investigation of the algorithmic properties of the model independent of its implementation details (Marr, 1980). It also, importantly, allows us to test our model on real-world data and compare directly against standard machine learning methods. Using actual thalamocortical circuitry to perform the unsupervised data clustering and the mechanism for the basal ganglia to provide reinforcement feedback, would be an interesting task for the distinct goal of investigating potential implementation-level predictions; this holds substantial potential for future research.

We emphasize that our focus is to use existing hypotheses of telencephalic component function already posited in the literature; these mechanisms lead us to specifically propose a novel method by which supervised learning is achieved by the unlikely route of combining unsupervised learning with reinforcement. This kind of computational-level abstraction and analysis of biological entities continues in the tradition of many prior works, including Suri and Schultz (2001), Schultz (2002), Daw and Doya (2006), Lee and Mumford (2003), Rodriguez et al. (2004), George and Hawkins (2009), Marr (1980), Riesenhuber and Poggio (1999), Ullman (2006), and many others.

## 3. SIMPLIFIED ALGORITHM

In our simplified algorithm, we refer to a method which we term *PARTITION*, corresponding to any of a family of clustering methods, intended to capture the clustering functionality of thalamocortical loops as described in the previous sections; and we refer to a method we term *SUBDIVIDE*, corresponding to any of a family of simple reinforcement methods, intended to capture the reinforcement-learning functionality of the basal ganglia/striatal complex as described in the previous sections. These operate together in an iterative loop corresponding to cortico-striatial (cluster–reinforcement) interaction: *SUBDIVIDE* checks for the "terminating" conditions of the iterative loop by examining the labels of the constituents of a given cluster and returning a *true* or *false* response. The resulting training method builds a tree of categories which, as will be seen, has the effect of performing supervised learning of the classes. The leaves of the tree contain class labels; the intermediate nodes may contain members of classes with different labels. During testing, the tree is traversed to obtain the label prediction for the new samples. Each data sample (belonging to one of $K$ labeled classes) is represented as a vector $x \in R^m$. During training, each such vector $x_i$ has a corresponding label $y_i \in 1, \ldots K$. (The subsequent "Experiments" section below describes the methods used to transform raw data such as natural images into vector representations in a domain-dependent fashion.)

### 3.1. TRAINING

The input to the training procedure is the training dataset consisting of $\langle x_i, y_i \rangle$ pairs where $x_i$ is an input vector and $y_i$ is its intended class label, as in all supervised learning methods. The output is a tree that is built by performing a succession of unsupervised splits of the data. The data corresponding to any given node in the tree is a subset of the original training dataset with the full dataset corresponding to the root of the tree. The action performed with the data at a node in the tree is an unsupervised split, thereby generating similarity-based clusters (subclusters) of the data within that tree node. The unsupervised split results in expansion (deepening) of the tree at that node, with the child nodes corresponding to the newly created unsupervised data clusters. The cluster representations corresponding to the children are recorded in the current node. These representations are used to determine the local branch that will be taken from this node during testing, in order to obtain a class prediction on a new sample. For each of the new children nodes, the labels of the samples within the cluster are examined, and if they are deemed to be sufficiently pure, i.e., a sufficient percentage of the data belong to the same class, then the child node becomes a (terminal) leaf in the tree. If not, the node is added to a queue which will be subjected to further processing, growing the tree. This queue is initialized with the root of the tree. The procedure (sketched in **Algorithm 1** below) proceeds until the queue becomes empty.

To summarize the mechanism, the algorithm attempts to find clusters based on appearance similarity, and when these clusters don't match with the intended (supervised) categories, reinforcement simply gives the algorithm the binary command to either split or not split the errant cluster. The behavior of the algorithm on sample data is illustrated in **Figure 2**. The input space of images is partitioned by successively splitting the corresponding training samples into subclusters at each step.

#### 3.1.1. *Picking the right branch factor*

Since the main free parameter in the algorithm is the number of unsupervised clusters to be spawned from any given node in the hierarchy, the impact of that parameter on the performance of the algorithm should be studied. This quantity corresponds to the branching factor for the class tree. We initially propose a single parameter as an upper bound for the branch factor: $K^{max}$, which fixes the largest number of branches that can be spawned from any node in the tree. Through experimentation (discussed in the Results section) we have determined that (i) very small values for this parameter result in slightly lower prediction accuracy;

```
Input: Dataset: X = {x_i ∈ R^M} with labels
       Y = {y_i ∈ {1, 2, 3..K}}
Output: Class Tree: A tree rooted at the node
       TRoot
Init: TRoot.X = X, TRoot.Y = Y; TRoot.Labels =
LABELSET(Y)
Q = []; Add(Q, TRoot);
while Q is not empty do
  qn = First node in Q
  if SUBDIVIDE(X_qn, Y_qn) = true then
    [Centroids, Clusters] = PARTITION (X_qn,
    K*)
    foreach Cluster C_k do
      Node T
      T.X = Clusters[k]
      T.Labels = LABELSET(Y(T.X))
      qn.Branches[k] = Centroids[k]
      qn.Children[k] = T
      Add(Q, T)
    end
  end
end
```

**Algorithm 1 |** A sketch of the CSL learning algorithm. The method constructs a meta class tree, which records unsupervised structure within the data, as well as providing a means to perform class prediction on novel samples. The function termed *PARTITION* denotes an unsupervised clustering algorithm, which can in principle be any of a family of clustering routines; selections for this algorithm are described later in the text. The subroutine *SUBDIVIDE* determines if the data at a tree node qn all belong to the same class or not. If the data come from multiple classes, *SUBDIVIDE* returns *true* and otherwise, *false*. See text for further description of the derivation of the algorithm.

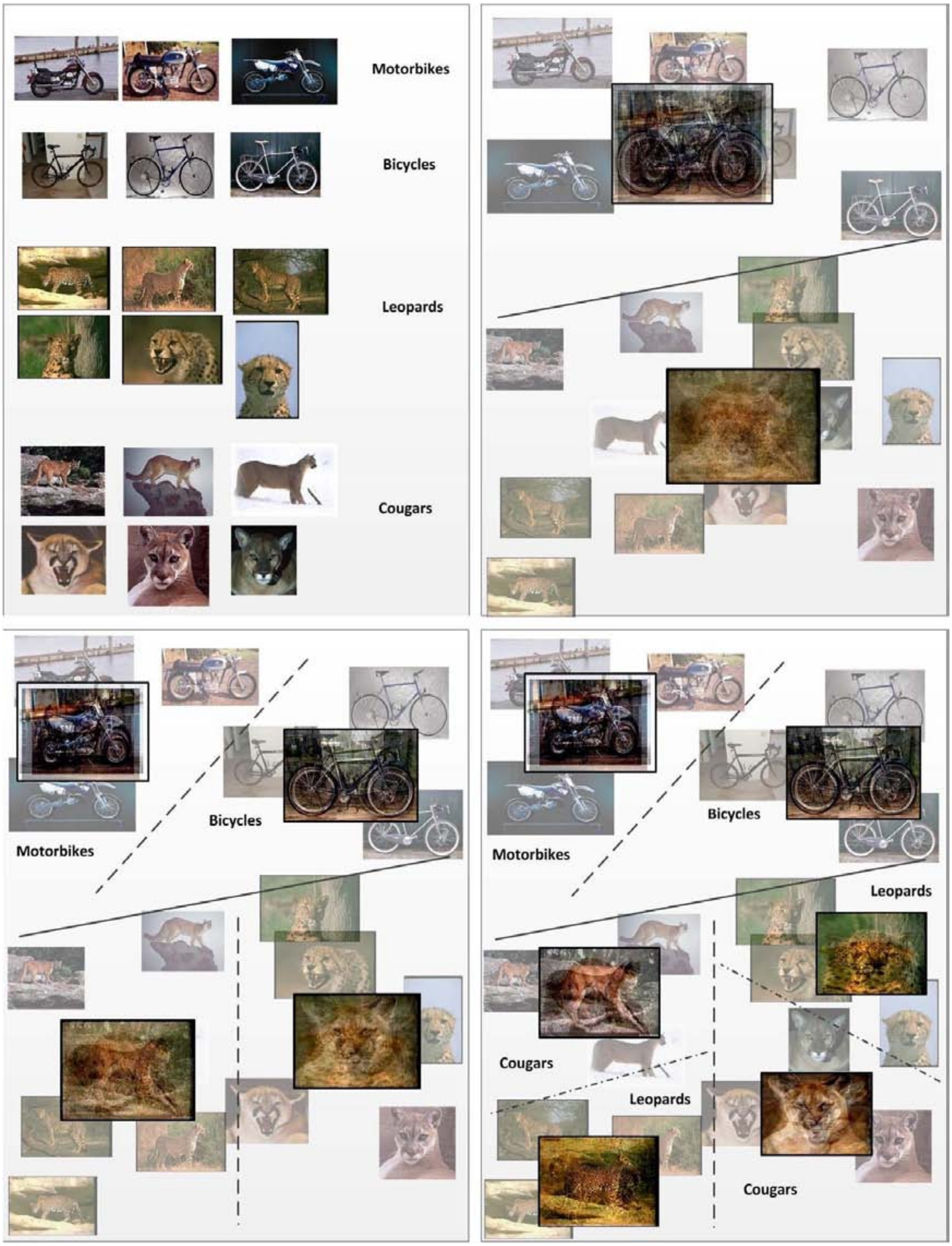

**FIGURE 2 | A simplified illustration of the iterative learning process with labeled data.** Images are successively split into two partitions in an unsupervised fashion (i.e., by similarity). The partitioning of data proceeds iteratively until the clusters formed are pure with respect to their labels. For each split, on either side of the dividing hyperplane, the means are shown as an overlay of the images that fall on the corresponding side of the hyperplane.

(ii) for sufficiently large values, the parameter setting has no significant impact on the performance efficacy of the classifier; and (iii) larger values of the parameter modestly increase the memory requirements of the clustering algorithm and thus the runtime of the learning stage (see the Results section below for further detail). (It is worth noting that selection of the best branch factor value may be obtained by examination of the distribution of the data to be partitioned in the input space, enabling automatic selection of the ideal number of unsupervised clusters without reference to the number of distinct labeled classes that occur in the space. Future work may entail the study of existing methods for this approach, Baron and Cover, 1991; Teh et al., 2004, as potential adjunct improvements to the CSL method.)

### 3.2. TREE PRUNING

Categorization algorithms are often subject to overfitting the data. Aspects of the CSL algorithm can be formally compared to those of decision trees, which are subject to overfitting.

Unlike decision trees, the classes represented at the leaves of the CSL tree need not be regarded as conjunctions of attribute values on the path from the root, and can be treated as fully represented classes by themselves. (We refer to this as the "leaf independence" property of the tree; this property will be used when we describe testing of the algorithm in the next section.) Also, since the splits are unsupervised and based on multidimensional similarity (also unlike decision trees), they exhibit robustness w.r.t. variances in small subsets of features within a class.

Both of these characteristics (leaf independence and unsupervised splitting) theoretically lead to predictions of less overfitting of the method.

In addition to these formal observations, we studied overfitting in the CSL method empirically. Analogously to decision trees, we could choose either to stop growing the tree before all leaves were perfectly pure (and potentially overfit), or to build a full tree and then somewhat prune it back. Both methods improve the overfitting problem observed in decision trees. Experiments with both methods in the CSL algorithm found that neither one had a significant effect on prediction accuracy. Thus, surprisingly, both theoretical and empirical studies find that the CSL class trees generalize well without overfitting; the method is unexpectedly resistant to overfitting.

### 3.3. TESTING

During testing, the algorithm is presented with previously unseen data samples whose class we wish to predict. The training phase created an appearance-based class hierarchy. Since the tree, including the "pure class" leaves, is generative in nature, there are two alternative procedures for class prediction. One is that of descending the tree, as is done in decision trees. However, in addition, the "leaf independence" property of the CSL tree, as described in the previous section (which does not hold for decision trees), enables another testing method, which we refer to as KNN-on-leaves, in which we only attend to the leaf nodes of the tree, as described in the second subsection below. (This property does not hold for decision trees, and thus this additional testing method cannot be applied to decision trees). The two test methods have somewhat different memory and computation costs and slightly different prediction accuracies.

#### *3.3.1. Tree descent*

This approach starts at the root of the class tree, and descends. At every node, the test datum is compared to the cluster centroids stored at the node to determine the branch to take. The branch taken corresponds to the closest centroid to the test datum; i.e., a decision is made locally at the node. This provides us a unique path from the root of the class hierarchy to a single leaf; the stored category label at that leaf is used to predict the label of the input. Due to tree pruning (described above), the leaves may not be completely pure. As a result, instead of relying on any given class being present in the leaves, the posterior probabilities for all the categories represented at the leaf are used to predict the class label for the sample.

#### *3.3.2. KNN-on-leaves*

In this approach, we make a note of all the leaves in the tree, along with the cluster representation in the parent of the leaf node corresponding to the branch which leads to the leaf. We then do K-nearest neighbor matching of the test sample with all these cluster centroids that correspond to the leaves. The final label predicted corresponds to the label of the leaf with the closest centroid. This approach implies that only the leaves of the tree need to be stored, resulting in a significant reduction in the memory required to store the learned model. However, a penalty is paid in recognition time, which in this case is proportional to the number of leaves in the tree.

The memory required to store the model in the tree descent approach is higher than that for the KNN-on-leaves approach. However, tree descent offers a substantial speedup in recognition, as comparisons need to be performed only along a single path through the tree from the root to the final leaf. The algorithm is sketched below in **Algorithm 2**.

We expect that the KNN-on-leaves variant will yield better prediction accuracy as the decision is made at the end of the tree and

**Input**: $x \in R^M$ , Class tree: TRoot
**Output**: $y \in 1, 2, ...K$
**Init:** Tree Node T = TRoot
**while** *T is not leaf* **do**
  $mostSim = 0$
  **for** $k$ = 1: $|T.Children|$ **do**
    sim = SIMILARITY(x, T.Centroids[k])
    **if** $sim > mostSim$ **then**
      mostSim = sim
      branch = k
    **end**
  **end**
  T = T.Children[branch]
**end**
y = T.LabelSet

**Algorithm 2 |** A sketch of the tree descent algorithm for classifying a new data sample. The method starts at the root node and descends, testing the sample datum against each node encountered to determine the branch to select for further descent. The result is a unique path from the root to a single leaf; the stored category at that leaf is the prediction of the label of the input. In the event of impure leaves, the posterior probabilities for all categories in the leaf are used to predict the class label of the sample. See text for further description.

hence the partitioning of the input space is expected to exhibit better generalization. In the case of tree descent, since decisions are made locally within the tree, if the dataset has high variance, then it is possible that a wrong branch will be taken early on in the tree, leading to inaccurate prediction. This problem is common to a large family of algorithms, including decision trees. We have performed experiments to compare the two test methods; the results confirm that the KNN-on-leaves method exhibits marginally better prediction than the tree-descent method. The behavior of the two methods is illustrated in **Figure 3**.

## 4. CLUSTERING METHODS

The only remaining design choice is which unsupervised clustering algorithm to employ for successively partitioning the data during training, and the corresponding similarity measure. The choice can change depending on the type of data to be classified, while the overall framework remains the same, yielding a potential family of closely related variants of the CSL algorithm. This enables flexibility in selecting a particular unsupervised clustering algorithm for a given domain and dataset, without modifying anything else in the algorithm. (Using different clustering algorithms within the same class tree is also feasible as all decisions are made locally in the tree.)

There are numerous clustering algorithms from the simple and efficient k-means (Lloyd, 1982), self organizing maps (SOM; Kaski, 1997) and competitive networks (Kosko, 1991), to the more elaborate and expensive probabilistic generative algorithms like mixture of Gaussians, Probabilistic latent semantic analysis (PLSA; Hoffman, 1999) and Latent Dirichlet Allocation (LDA; Blei et al., 2003); each has merits and costs. Given the biological derivation of the system, we began by choosing k-means, a simple and inexpensive clustering method that has been discussed previously as a candidate system for biological clustering (Darken and Moody, 1990); the method could instead use SOM or competitive learning, two highly related systems. (It remains quite possible that more robust (and expensive) algorithms such as PLSA and LDA could provide improved prediction accuracy. Improvements might also arise by treating the data at every node as a mixture of Gaussians, and estimating the mixture parameters using the expectation maximization (EM) algorithm.)

### 4.1. *k*-MEANS

*k-Means* is one of the most popular algorithms to cluster $n$ vectors based on distance measure into $k$ partitions, where $k < n$. It attempts to find the centers of natural clusters in the data. The objective that $k$-means tries to minimize is the total *intra cluster*

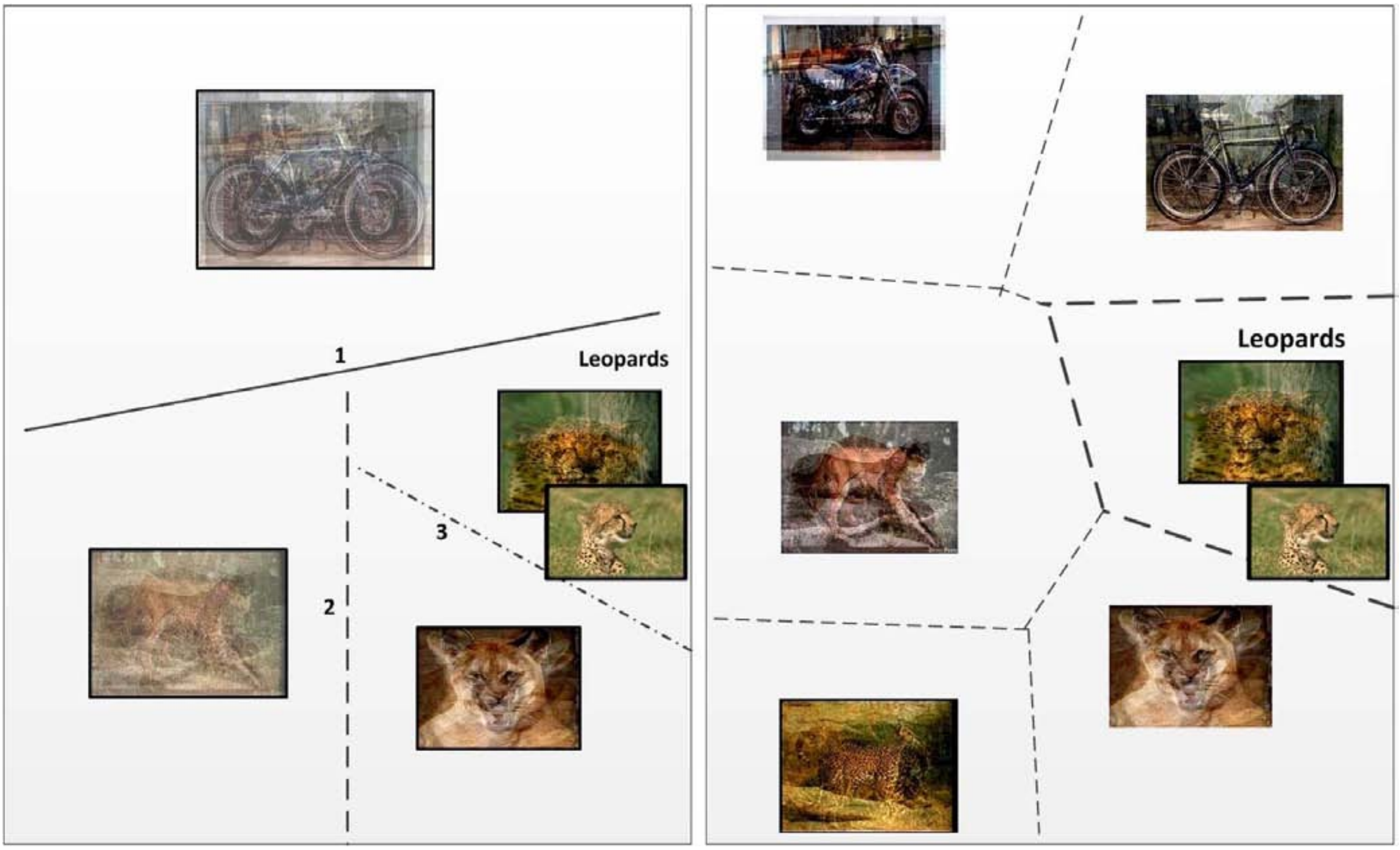

**FIGURE 3 | Two methods by which the CSL algorithm predicts category membership at test time.** (Left) Class prediction via hierarchical descent. At each step, a new (unknown) sample will fall on one or the other side of a classification hyperplane. The decision provides a path through the class tree at each node. At the leaves, the class prediction is obtained. The numbering gives the order in which the hyperplanes are probed. (Right) Class prediction using only leaves of class tree. All leaves are considered simultaneously; the test sample is compared to each leaf and the class prediction is obtained using KNN.

*variance*, or, the squared error function:

$$\Phi = \sum_{i=1}^{K} \sum_{x_j \in C_i} (x_j - \mu_i)^2$$

where there are $K$ clusters $S_i$, $i = 1, 2, \ldots, K$ and $\mu_i$ is the centroid or mean point of all the points $x_j \in C_i$.

When k-means is used for the unsupervised appearance-based clustering at the nodes of the class tree, the actual means obtained are stored at each node, and the similarity measure is inversely proportional to the Euclidean distance.

#### 4.1.1. Initializing clusters

In general, unsupervised methods are sensitive to initialization. We initialize the clustering algorithm at every node in the class tree as follows.

If we are at node $l$, with samples having one of $K_l$ labels, we first determine the class averages of the $K_l$ categories. (For every class, we remove the samples which are at least 2 standard deviations away from the mean of the class for the initialization These samples are considered for the subsequent unsupervised clustering.) If the number of clusters (branches), $K^* = min\ (K_l, K^{max})$ turns out to be equal to $K_l$, then the averages are used as the seeds for the clustering algorithm. If however $K^* < K_l$, then we use a simple and efficient method for obtaining the initial clusters by using an initial run of k-means on the $K_l$ averages in order to obtain the $K^*$ initial centroids. The data samples are assigned to the clusters using nearest neighbor mapping, and the averages of these $K^*$ clusters are used as seeds for a subsequent run of the unsupervised clustering algorithm. (In our empirical experiments we have used the k-means++ variant of the popular clustering algorithm to obtain the initial cluster seeds; Arthur and Vassilvitskii, 2007.) **Figure 4** illustrates the initialization method. (While the method works relatively well, further studies indicate that other methods, which directly utilize the semantic structure of the labeled dataset, can result in even better performance. These alternate approaches are not discussed in this paper in order to keep the focus on introducing the core algorithm.) It is worth noting that the initialization method can be thought of in terms of a logically prior "developmental" period, in which no data is actually stored, but instead sampling of the environment is used to set parameters of the method; those parameters, once fixed, are then used in the subsequent performance of the then-"adult" algorithm (Felch and Granger, 2008).

## 5. EXPERIMENTS

The proposed algorithm performs a number of operations on its input, including the unsupervised discovery of structure in the data. However, since the method, despite being composed only of unsupervised clustering and reinforcement learning, can nonetheless perform supervised learning, we have run tests that involve using the CSL method solely as a supervised classifier. In addition to these tests of supervised learning alone, we then briefly describe some additional findings illustrating the CSL algorithm's power at tasks beyond the classification task (including the tasks of identifying structure in data, and localizing objects within images).

When viewed solely as a supervised classifier, the CSL method bears resemblances to two well-studied methods in machine learning and statistics, and we rigorously compare these. We compared the accuracy, and the time and space costs, of the CSL algorithm as a supervised classifier, against the support vector machine (SVM) and k-nearest neighbor (KNN) algorithms. Performance was examined on two well-studied public datasets.

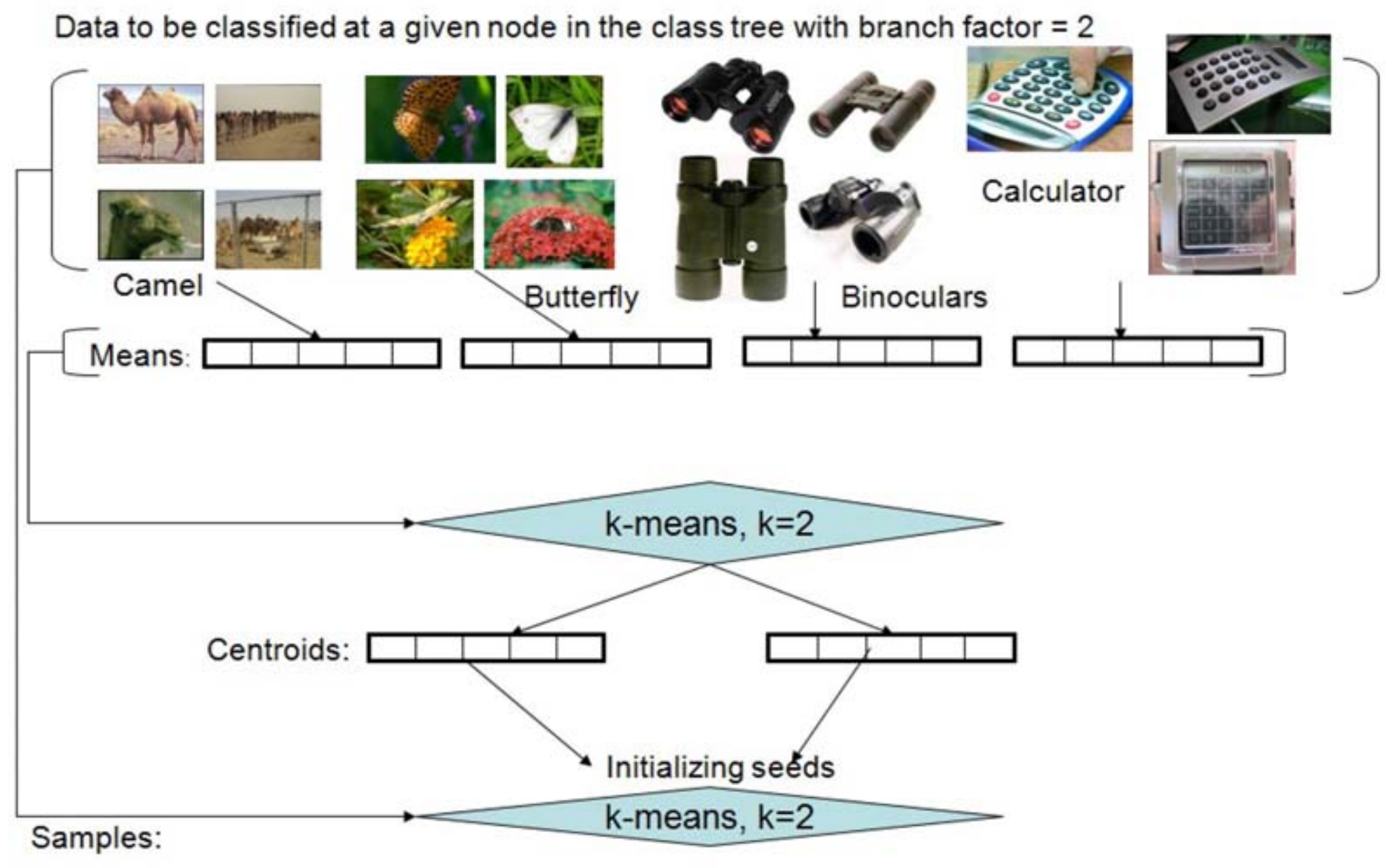

**FIGURE 4 | Initialization of the unsupervised partitioning for a set of labeled training examples.** The illustration depicts the process when working with image datasets. The method needs to be applied only when the desired number of clusters (in this example, 2) is less than the actual number of labeled classes represented in the dataset (in this example, 4) to be clustered.

For SVM, we have used the popular LibSVM implementation that is publicly available (Chang and Lin, 2001). This package implements the "one vs one" flavor of multiclass classification, rather than "one vs rest" variant based on the findings reported in Hsu and Lin (2002). After experimenting with a few kernels, we chose the linear kernel since it was the most efficient and especially since it provided the best SVM results for the high-dimensional datasets we tested. It is known that for the linear kernel a weight vector can be computed and hence the support vectors need not be kept in memory, resulting in low memory requirements and fast recognition time. However, this is not true for non-linear kernels where support vectors need to be kept in memory to get the class predictions at run time. Since we wish to compare the classifiers in the general setting and it is likely that the kernel trick may need to be employed to separate non-linear input space, we have retained the implementation of LibSVM as it is (where the support vectors are retained in memory and used during testing to get class prediction). We realize this may not be the fairest comparison for the current set of experiments, however, we believe that this setting is more reflective of the typical use case scenario where the algorithms will be employed.

For KNN we have hand coded the implementation and set the parameter $K = 1$ for maximum efficiency. (For the CSL algorithm with KNN-on-leaves, we use $K = 1$ as well.) The test bed is a machine running windows XP 64 with 8GB memory. We have not used hardware acceleration for any of the algorithms to keep the comparison fair.

We have used two popular datasets from different domains with very different characteristics (including dimensionality of the data) to fully explore the strengths and weaknesses of the algorithm. One is a subset of the Caltech-256 image set, and the other is a very high-dimensional dataset of neuroimaging data from fMRI experiments, that has been widely studied.

For both experiments, we performed multiple runs, differently splitting the samples from each class into training and testing sets (roughly equal in number). The results shown indicate the means and standard deviations of all runs.

## 5.1. OBJECT RECOGNITION

Our first experiment tests the algorithm for object recognition in natural still image datasets. The task is to predict the label for an image, having learned the various classes of objects in images through a training phase. We report empirical findings for prediction accuracy and computational resources required.

### 5.1.1. Dataset

The dataset used consists of a subset of the Caltech-256 dataset (Griffin et al., 2007) using 39 categories, each with roughly 100 instances. The categories were specifically chosen to exhibit very high between-category similarity, intentionally selected as a very challenging task, with high potential confusion among classes. The categories are:

- Mammals: bear, chimp, dog, elephant, goat, gorilla, kangaroo, leopard, raccoon, zebra
- Winged: duck, goose, hummingbird, ostrich, owl, penguin, swan, bat, cormorant, butterfly
- Crawlers (reptiles/insects/arthropods/amphibians): iguana, cockroach, grasshopper, housefly, praying mantis, scorpion, snail, spider, toad
- Inanimate objects: backpack, baseball glove, binoculars, bulldozer, chandeliers, computer monitor, grand piano, ipod, laptop, microwave.

We have chosen an extremely simple (and very standard) method for representing images in order to maintain focus on the description of the proposed classifier. First a feature vocabulary consisting of SIFT features (Lowe, 2004) is constructed by running k-means on a random set of images containing examples from all classes of interest; each image is then represented as a histogram of these features. The positions of the features and their geometry is ignored, simplifying the process and reducing computational costs. Thus each image is a vector $x \in R^m$, where $m$ is the size of the acquired vocabulary. Each dimension of the vector is a count of the number of times the particular feature occurred in the image. This representation, known as the "Bag of Words," has been successfully applied before in several domains including object recognition in images (Sivic and Zisserman, 2003).

We ran a total of 8 trials, corresponding to 8 different random partitionings of the Caltech-256 data into training and testing sets. In each trial, we ran the test for each of a range of $K_{max}$ values, to test this free parameter of the CSL model.

### 5.1.2. Prediction accuracy

The graph in top left of **Figure 5** compares the classifier prediction accuracy of the proposed algorithm with that of SVMs on the 39 subsets of Caltech-256 described earlier. As expected, the simplistic image representation scheme, and the readily confused category members, renders the task extremely difficult. It will be seen that all classifiers perform at a very modest success rate with this data, indicating the difficulty of the dataset and the considerable room for potential improvement in classification techniques.

The two variants of the CSL algorithm are competitive with SVM: SVM has an average accuracy of 23.9%; CSL with tree descent has an average accuracy of 19.4%; and CSL with KNN-on-leaves has an average prediction accuracy of 21.3%. The KNN algorithm alone performs relatively poorly, with an average prediction accuracy of 13.6%. Chance probability of correctly predicting a class is 1 out of 39 (2.56%).

It can be seen that the branch factor does not have a significant impact on error rates. This is possibly because the class tree grows until the leaves are pure, and the resulting internal structure, though different across choices of $K^{max}$, does not significantly impact the ultimate classifier performance as the hierarchy adapts its shape. Different internal structure could significantly affect the performance of the algorithm on tasks that depended on the similarity structure of the data, but for the sole task of supervised classification, the tree's internal nodes have little effect on prediction accuracy.

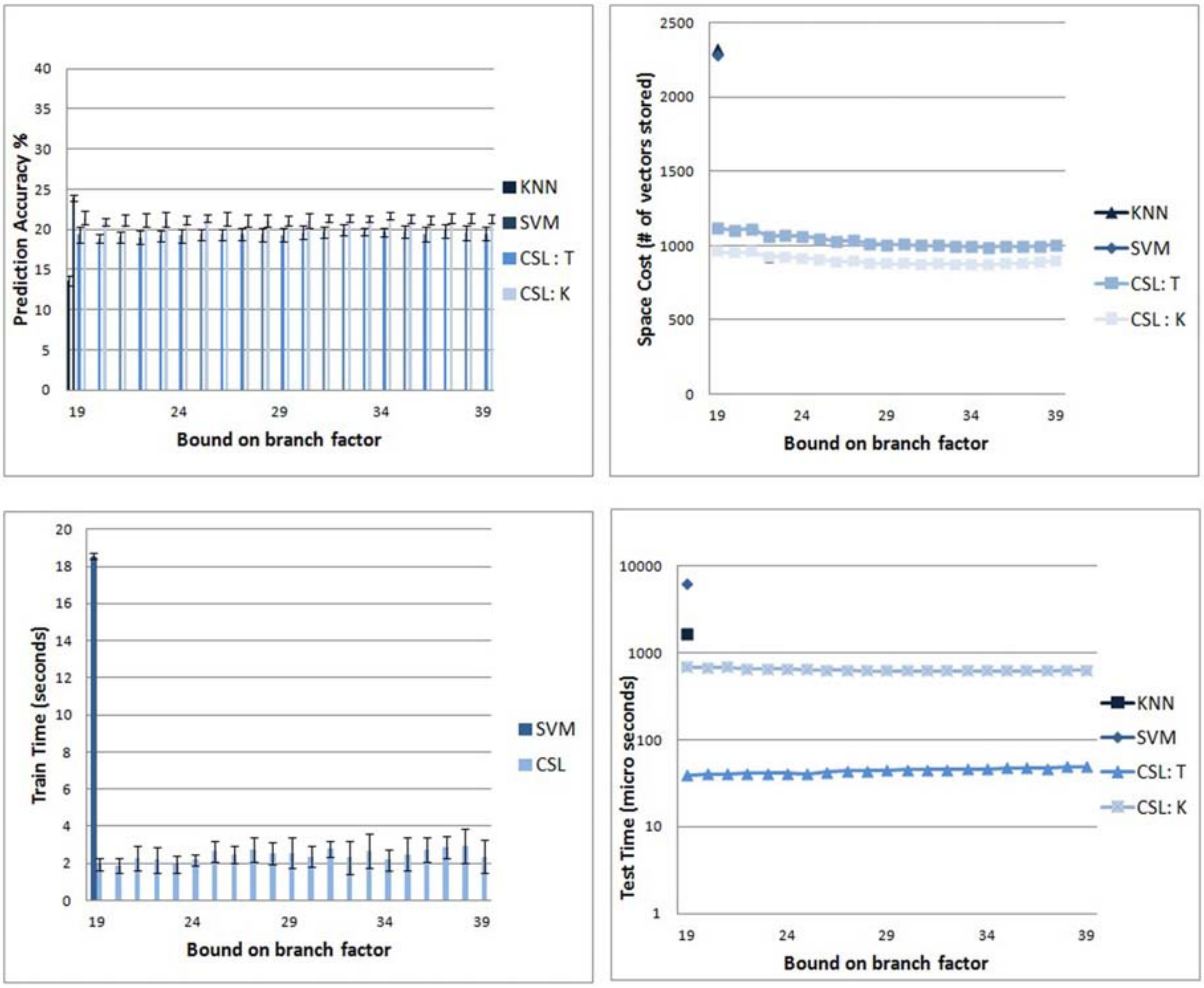

**FIGURE 5 | Comparison of CSL and other standard classifiers.** (Top left) Prediction accuracy of the CSL classifier on the Caltech-256 subsets. The scores in blue are the rates achieved by the CSL classifier. Scores in pink are from standard multiclass SVM (see text). (Top right) Memory requirements for CSL with respect to the branch factor parameter. The figure shows that the parameter does not significantly impact the size of the tree. Also, we can see a clear difference between the memory usage of CSL and the other supervised classifiers after training. (Bottom left) Run times for the training stage. CSL (red) requires roughly an order of magnitude less training run time than SVM (blue). (Bottom right) Average time to recognize a new image after training for the different algorithms. The y axis (logarithmic scale) shows CSL outperforming SVM and KNN by an order of magnitude.

#### *5.1.3. Memory usage*

The graph in top right of **Figure 5** shows the relationship between the overall number of nodes in the tree to be retained (and hence vectors of dimensionality $M$) and the branch factor for CSL classifier. CSL with tree descent had to store an average of 1036.25 vectors, while the knn-on-leaves variant had to store 902.21 vectors. SVM required 2286 vectors while the vanilla KNN method (with $k = 1$) requires storage of the entire training corpus of 2322 vectors. Thus, the number of vectors retained in memory by the CSL variants is roughly half the number retained by the SVM and KNN algorithms. Further, the memory needed to store the trained model when we predict using the KNN-on-leaves approach is smaller than when we use tree descent, as we expected and discussed earlier. As can be seen, there is not much variation in CSL performance across different branch factor values. This suggests that after a few initial splits, most of the sub trees have very few categories represented within them and hence the upper bound on the branch factor does not play a significant role in ongoing performance.

#### *5.1.4. Classifier run times*

The runtime costs of the algorithms paint an even more startling picture. The graph in bottom left of **Figure 5** shows the plots comparing the training times of the CSL and SVM algorithms. The two variants of CSL have the same training procedure and hence require the same time to train. (KNN has no explicit training stage.) As can be seen, the training time of the new algorithm (average of 2.42 s) is roughly an order of magnitude smaller than that of the SVM (average of 18.54 s). It should be clearly noted that comparisons between implementations of algorithms will not necessarily reflect underlying computational costs inherent to the algorithms, for which further analysis and formal treatment will be required. Nonetheless, in the present experiments, the empirical costs were radically different despite efforts to show the SVM in its best light.

As indicated earlier, the choice of branch factor does not have a large impact on the training time needed. We also found that the working memory requirements of our algorithm were very small compared to that of the SVM. In the extreme, when large representations were used for images, the memory requirements for SVMs rendered the task entirely impracticable. In such circumstances, the CSL method still performed effectively. The working amount of memory we need is proportional to the largest clustering job that needs to be performed. By choosing low values of $K^{max}$, we empirically find that we can keep this requirement low without loss of classifier performance.

The bottom right plot of **Figure 5** shows how the average time for recognizing a new image varies with branch factor. The times are shown in logarithmic scale. The CSL variants are an order of magnitude faster than KNN and SVM algorithms with the tree descent variant being the fastest. This shows the proposed algorithm in its best light. Once training is complete, recognition can be extremely rapid by doing hierarchical descent, making the CSL method unusually well suited for real-time applications.

## 5.2. HAXBY fMRI DATASET, 2001

### *5.2.1. Dataset*

Having demonstrated the CSL system on image data, we selected a very different dataset to test: neuroimaging data collected from the brain activity of human subjects who were viewing pictures. As with the Caltech-256 data, we selected a very well-studied set of fMRI data, from a 2001 study by Haxby et al. (2001).

Six healthy human volunteers entered an fMRI neuroimaging apparatus and viewed a set of pictures while their brain activity (blood oxygen-level dependent measures) was recorded. In each run, the subjects passively viewed gray scale images of eight object categories, grouped in 24 s blocks separated by rest periods. Each image was shown for 500 ms and was followed by a 1500-ms interstimulus interval. Each subject carried out twelve of these runs. The stimuli viewed by the subjects consisted of images from the following eight classes: Faces, Cats, Chairs, Scissors, Houses, Bottles, Shoes, and random scrambled pictures. Full-brain fMRI data were recorded with a volume repetition time of 2.5 s, thus, a stimulus block was covered by roughly 9 volumes. For a complete description of the experimental design and fMRI acquisition parameters, see Haxby et al. (2001). (The dataset is publicly available.) Each fMRI recording corresponding to 1volume in a block for a given input image can be thought of as a vector with 163840 dimensions. The recordings for all the subjects have the same vector length. (In the original work, "masks" for individual brain areas were provided, retaining only those voxels that were hypothesized by the experimenters to play a significant role in object recognition. Using these masks reduces the data dimensionality by a large factor. However, the masks are of different lengths for different subjects, thus preventing meaningful aggregation of recordings across subjects. Thus, we have not used the masks and instead trained the classifiers in the original high dimensional space.)

### *5.2.2. Testing on individual subjects*

For each subject who participated in the experiment, we have neuroimaging data collected as that subject viewed images from each of the eight classes. The task was to see whether, from the brain data alone, the algorithms could predict what type of picture the subject was viewing. Top left in **Figure 6** shows the prediction accuracy of the various classifiers we tried. On the whole, all the classifiers exhibit similar performance with SVM performing slightly better on a couple of the subjects.

Top right of **Figure 6** shows the memory requirements for all the algorithms. The CSL variants require significantly less memory to store the model learned during training compared to SVM and KNN. SVM requires a large number of support vectors to fully differentiate the data from different classes leading to large memory consumption, whereas KNN needs to store all the training data in memory. For CSL, if the testing method is tree descent, then the entire hierarchy needs to be kept in memory. For the KNN-on-leaves testing method, only the leaves of the tree are retained, rendering even a smaller memory requirement for the stored model.

Bottom left of **Figure 6** shows the training time for the CSL algorithm being an order of magnitude smaller than that of SVM. KNN does not have any explicit training stage. Finally, bottom right of **Figure 6** compares the recognition time for the different algorithms, again on a log scale. The average recognition time on a new sample for the CSL tree descent variant is a couple of orders of magnitude smaller than both KNN and SVM. For the KNN-on-leaves variant of the CSL method, the recognition time grows larger (while still being significantly smaller than KNN or SVM). Therefore the fastest approach is performing a tree descent (paying a penalty in terms of memory requirements for storing the model).

### *5.2.3. Aggregating data across subjects*

Since the recordings from all the subjects have the same dimensionality, we can merge all the data from the different subjects into 1 large dataset and partition it into the training and testing datasets. This way we can study the performance trends with increasing datasets. The SVM system, unfortunately, was unable to run on pools containing more than two subjects, due to the SVM system's high memory requirements. Nonetheless, the two variants of the CSL algorithm, and the KNN algorithm, ran successfully on collections containing up to five subjects' aggregated data.

The subplot on the left of **Figure 7** shows that the classification prediction accuracy of the different classifiers remain competitive with each other as we increase the pool. The subplot on the right of **Figure 7** shows the trend of memory consumption by the different algorithms as we increase the number of subjects included. Compared to standard KNN, the increase in memory consumption is much slower (sub linear) for the CSL algorithm, with the KNN-on-leaves variant of the CSL algorithm growing very slowly.

Finally, in **Figure 8**, we examine the growth in the average recognition time with increasing pool size. The costs of adding data cause the recognition time to grow for the KNN algorithm more than for either variant of the CSL algorithm (either tree descent or KNN-on-leaves versions).

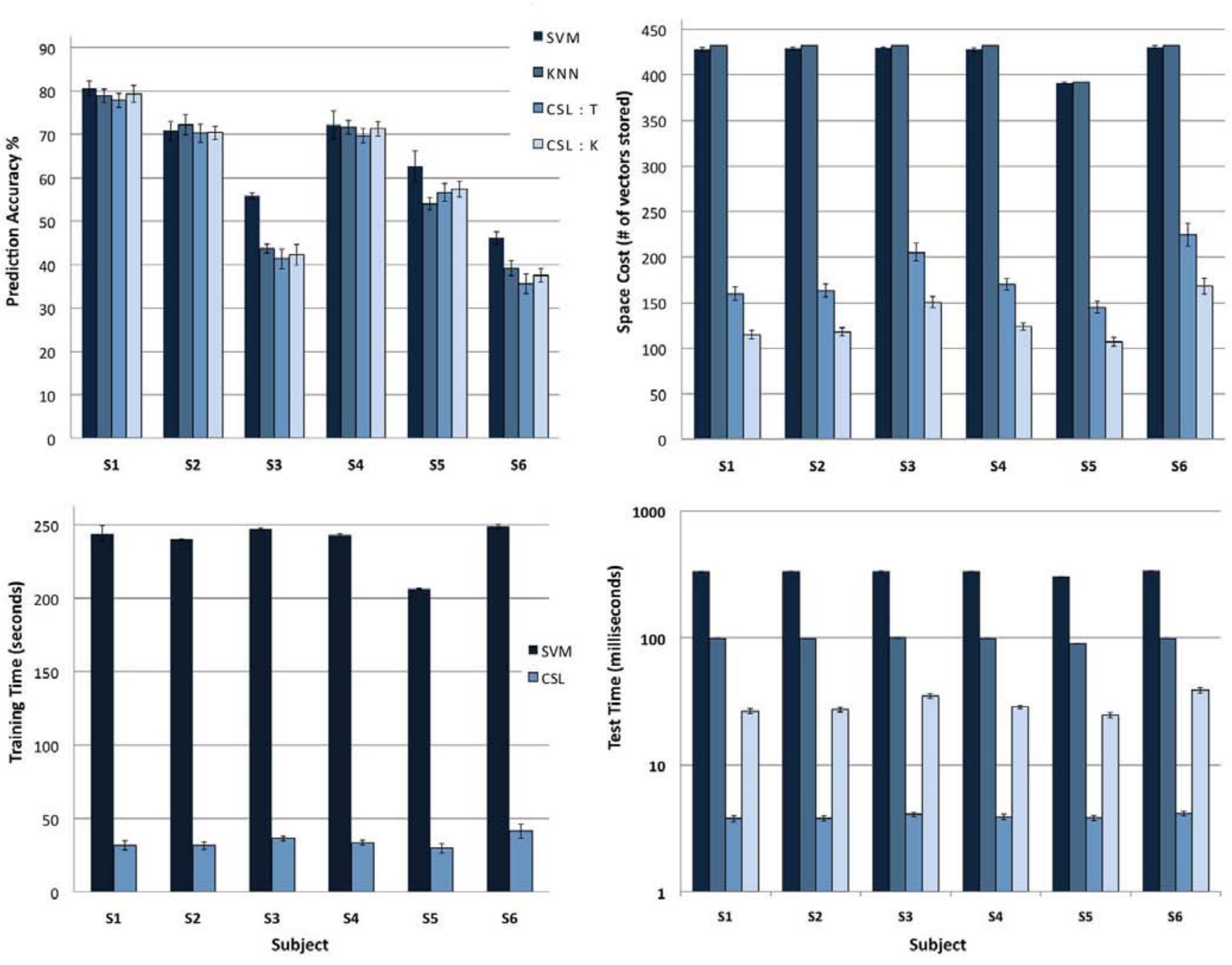

**FIGURE 6 | Comparison of classifiers.** The task was to determine, from neuroimaging data alone, what type of image the subject was viewing. The classifiers tested were SVM, KNN, CSL:KNN-on-leaves and CSL:Tree descent. Each test was run on the data from an individual subject in the fMRI experiment (see text). (Top left) Accuracy of classifier predictions. All classifiers have similar prediction accuracy for four of the subjects, while SVM performs slightly better on two subjects (S3, S6). (Top right) Memory usage across different algorithms. CSL variants exhibit significantly lower memory usage than the other two methods. (Bottom left) Time required to train classifiers (CSL vs. SVM; the two CSL variants are tested differently but trained identically; KNN algorithms are not trained). CSL requires roughly an order of magnitude less time to train than SVM. (Bottom right) Average time to recognize a new input after training. The y axis (a logarithmic scale) shows clearly that CSL with tree descent outperforms the other classifiers by an order of magnitude.

Between these two CSL algorithm variants, the latter exhibits some modest time growth as data is added, whereas the former (tree descent) version of CSL exhibits no significant increase in recognition time whatsoever as more data is added to the task. It is notable that the reason for this is that the tree depth has not increased with increasing size of the dataset; that is, as more data is added, the learned CSL tree arrives at the ability to successfully classify the data early on, and adding new data does not require the method to add more to the tree. Interestingly, the trees become better balanced as we increase the number of subjects, but their sizes do not increase. The results suggest that the CSL algorithm is better suited to scale to extremely large data sets than either of the competing standard SVM or KNN methods.

## 6. ANALYSES AND EXTENSIONS

### 6.1. ALGORITHM COMPLEXITY

When k-means is used for clustering, the time complexity for each partitioning is O(NtK), where N is the number of samples, K is the number of partitions and t is the number of iterations. If we fix $t$ to be a constant (by putting an upper limit on it), then each split takes $O(NK)$. Since we also put a bound on K ($K_{max}$), we can assume that each split is O(N). Further analysis is needed on the total number of paths and their contribution to runtime. The maximum amount of memory needed is for the first unsupervised partitioning. This is proportional to $O(NK)$. When we have small $K$, the amount of memory is directly proportional to the number of data elements being used in training.

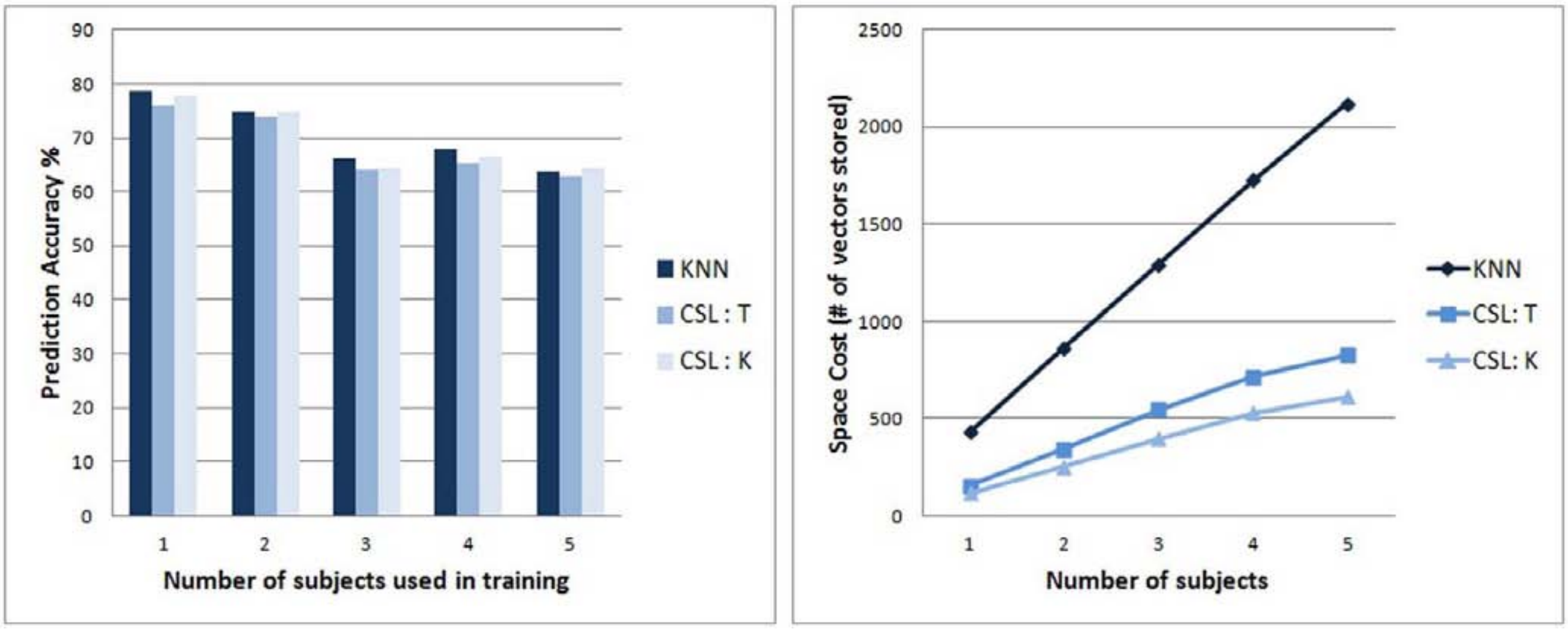

**FIGURE 7 | (Left) Accuracy of classifier predictions on split tests.** The accuracy of the different algorithms remain competitive as we increase the subject pool. (The memory usage by the LibSVM implementation of SVM was too large for testing on subject pools larger than 2). (Right) Tracks the trend of memory consumption with increasing size of the subject pool for the classifiers. KNN's memory usage grows linearly, whereas the CSL variants grow at a lower rate, illustrating their scalability.

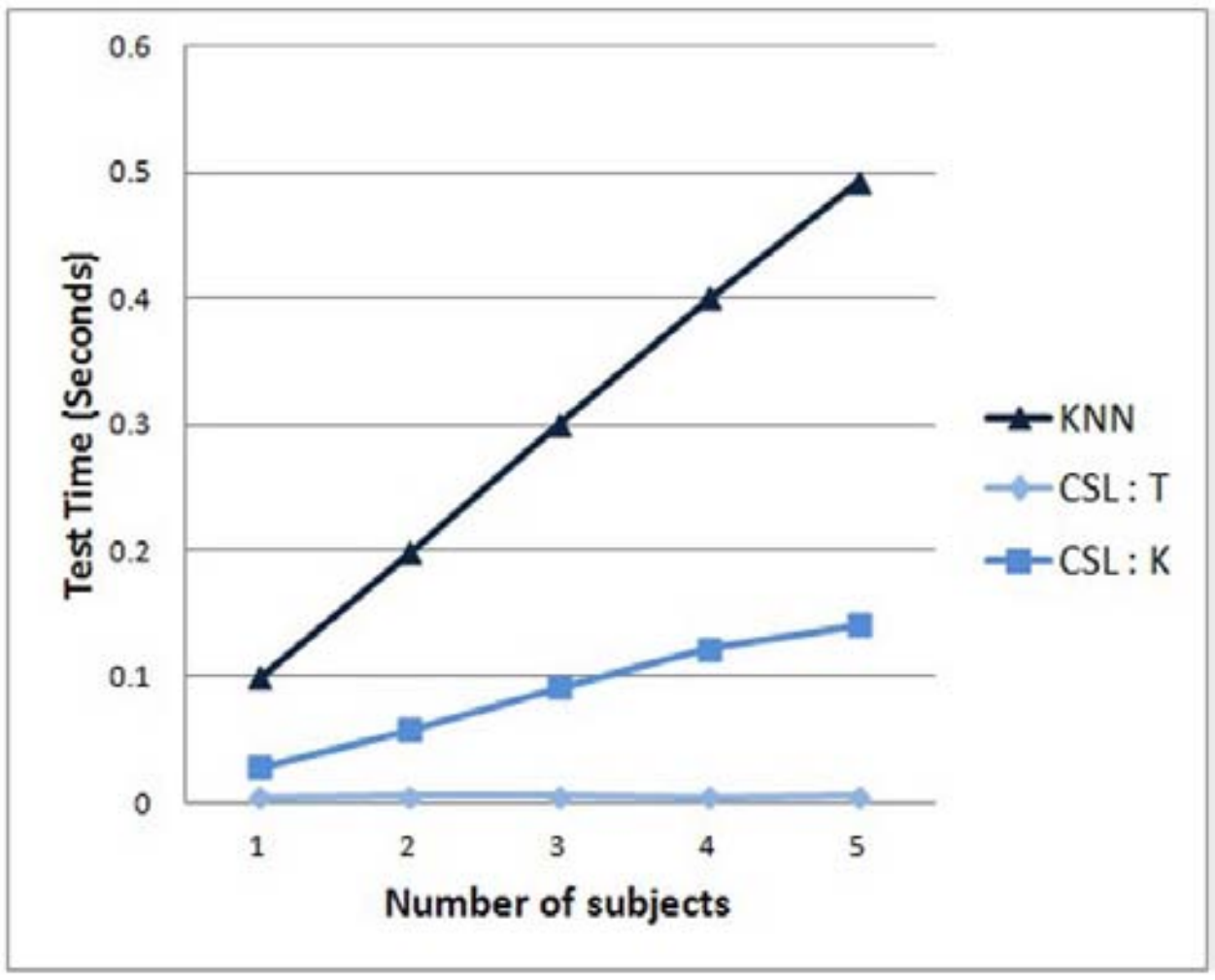

**FIGURE 8 | Average recognition time required for a new test sample, as a function of the amount of data trained.** As expected, KNN grows linearly. CSL with KNN-on-leaves grows more slowly. Interestingly, CSL with tree descent hardly shows any increase, suggesting its scalability to extremely large datasets.

As mentioned earlier, the algorithm is intrinsically highly parallel. After every unsupervised partitioning, each of the partitions can be further treated in parallel. However, in the experiments reported here, we have as yet made no attempt to parallelize the code, seeking instead to compare the algorithm directly against current standard SVM implementations.

### 6.2. COMPARISON WITH OTHER HIERARCHICAL LEARNING TECHNIQUES

The structure of the algorithm makes it very similar to CART (and in particular, decision trees; Buntine, 1992) since both families of algorithms partition the non-linear input space into discontinuous regions such that the individual sub regions themselves provide effective class boundaries. However, there are several significant differences.

- Perhaps the most substantial difference is that decision trees use the labels of the data to perform splits, whereas the CSL algorithm partitions based on unsupervised similarity.
- The CSL algorithm splits in a multivariate fashion, taking into account all the dimensions of the data samples, as opposed to decision trees where most often, a single dimension which results in the largest demixing of the data, is used to make splits. The path from the root to a leaf in a decision tree is a conjunction of local decisions on feature values and as a result is prone to over fitting. As discussed before, the CSL tends to exhibit little overfitting, and we can understand why this is the case (see Discussion in the Simplified Algorithm section earlier). The leaves can be treated independently of the rest of the tree and KNN can be used on them to obtain the class predictions.
- Decision trees are by nature a 2 class discriminative approach (multiclass problems can be handled using binary decision trees; Lee and Oh, 2003) whereas the CSL algorithm is a natural multiclass generative algorithm.

Most importantly, the goals of these systems differ. The primary goal of the CSL algorithm is to uncover natural structure within the data. The fact that the label-based impurity of classes is reduced, resulting in the ability to classify labeled data, falls out as a (very valuable) side effect of the procedure. The CSL algorithm thus will carry out a range of additional tasks, beyond supervised classification, that use deeper analysis of the underlying structure of the data, not apparent through supervised labeling alone.

### 6.3. DISCOVERY OF STRUCTURE

For purposes of this paper we have focused solely on the classification abilities of the algorithm, though the algorithm can perform many other tasks outside the purview of classification. Here we will briefly cover two illustrative additional abilities: (i) uncovering secondary structure of data, and (ii) localization of objects within images.

#### *6.3.1. Haxby dataset*

Once a model is trained, for each training sample if we do hierarchical descent and aggregate the posterior probabilities of the nodes along the path, we get a representation for the sample. When we do an agglomerative clustering on that representation, we uncover secondary structure suggesting meta classes occurring in the dataset. **Figure 9** captures the output of such an agglomerative clustering for the recordings of one subject (S1). Here we can see extensive structure relations among the responses to various pictures; perhaps most prominent is a clear separation of the data into animate and inanimate classes. The tree suggests the structure of information that is present in the neuroimaging data; the subjects' brain responses distinguish among the different types of pictures that they viewed. Related results were shown by Hanson et al. (2004); these were arrived at by analysis of the hidden node activity of a back propagation network trained on the same data. In contrast, it is worth noting that the CSL classifier obtains this structure as a natural byproduct of the tree-building process.

#### *6.3.2. Image localization*

A task quite outside the realm of supervised classification is that of localizing, i.e., finding an object of interest within an image. This task is useful to illustrate additional capabilities of the algorithm beyond just classification, making use of the internal representations it constructs.

We assume for this example that the clustering component of the algorithm is carried out by a generative method such as PLSA (Sivic et al., 2005); we then can assume that the features specific to the object class will contribute to the way in which an image becomes clustered, and that those features will contribute more than will random background features in the image.

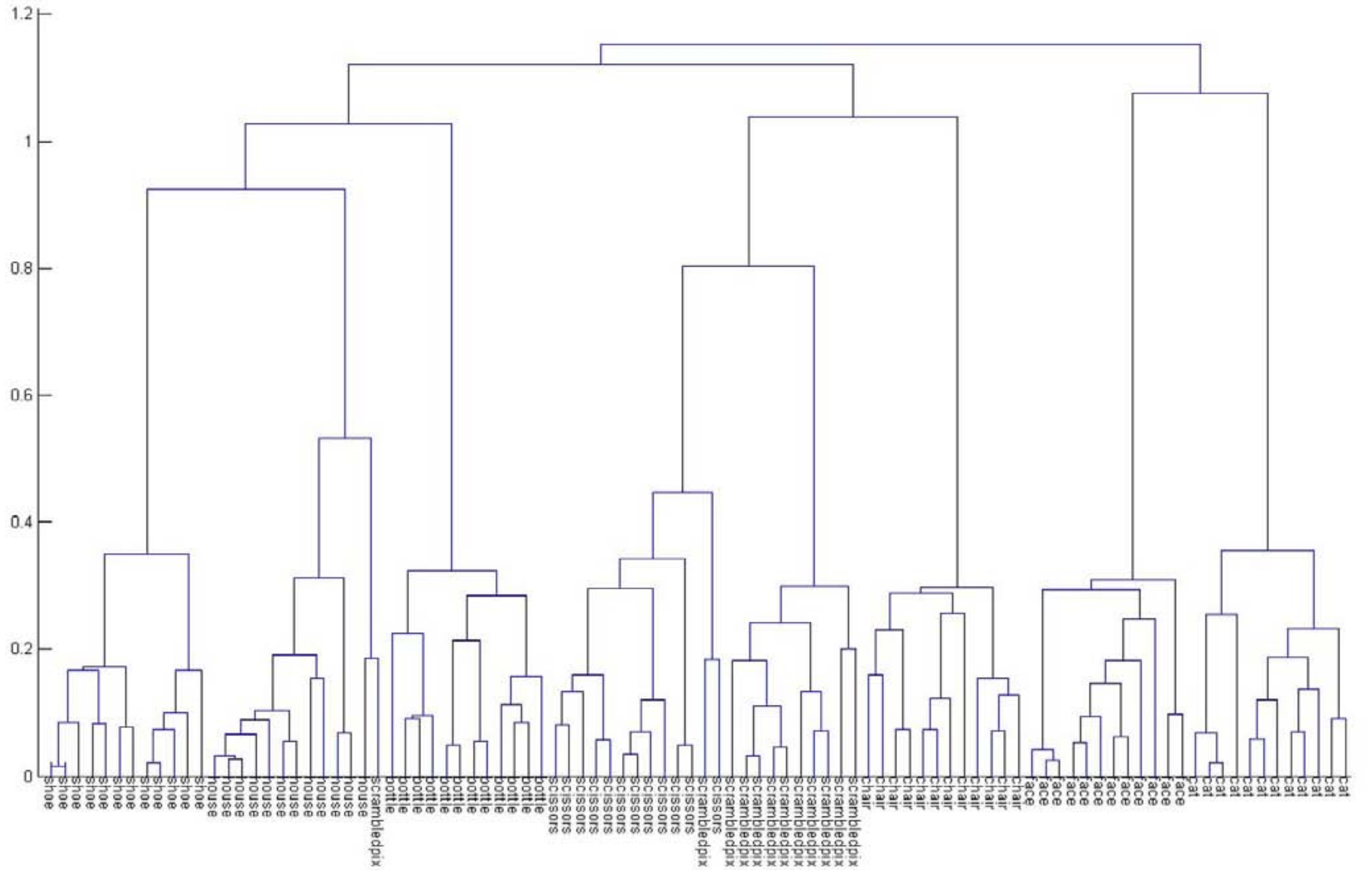

**FIGURE 9 | Agglomerative clustering of tree node activity for all the data from Subject 1.** The hierarchy shows extensive internal structure in the responses, including a clear separation of animate and inanimate stimuli. See text for further discussion.

**Figure 10** shows an example of localization of a face within an image. The initial task was to classify images of faces, cars, motorcycles, and airplanes (from Caltech 4). PLSA was used for clustering and for determining the cluster membership of a previously unseen image $x$. For every cluster $z$, we can obtain the posterior probability $p(z|x, w)$, for every feature $w$ in the vocabulary.

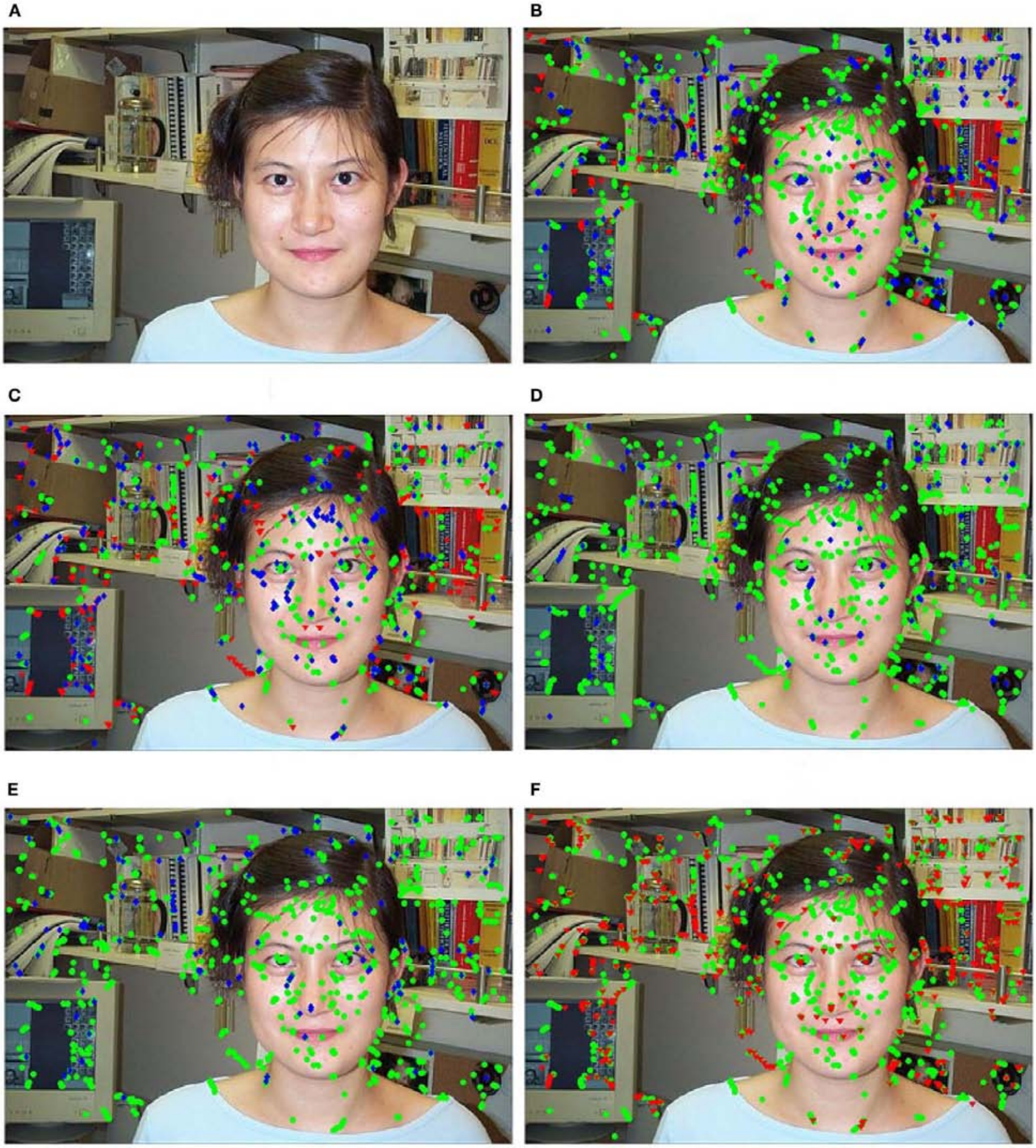

**FIGURE 10 | An illustration of object localization on an image from Caltech-256. (A)** The original image. **(B–E)** Positive, neutral, and negative features (green, blue, red, respectively) shown at levels 1 through 4 along the path in the CSL tree. **(F)** A thresholded map of the aggregate feature scores. Points in green indicate feature scores above threshold and those in red indicate below-threshold scores. Note that although green dots occur in multiple regions, the presence of red dots (negative features) is limited only to regions outside the face region.

Thus, we can test all features in the image to see which ones maximize the posterior, indicating strong influence on the eventual cluster membership. The location of those features can then be used to identify the vicinity of the object.

As the path from root to leaf in the CSL hierarchy is traversed for a particular test image, the posterior at a given node determines the contribution of the feature to the branch selected. Let $y$ be the final object label prediction for image $x$.

Consider feature $f_i$ from the vocabulary. At any given node at height $l$ along the path leading to prediction of $y$ for $x$, let $d_i^l$ be the branch predicted by feature $f_i$, i.e., among all branches at node $l$, the posterior for that branch is highest for that feature. $d_i^l$ is actually a set of labels that can be reached at various leaves using the branch and finally let the overall branch taken at $l$ be $b^l$.

At level $l$, $f_i$, can be classified as positive if $1(d_i^l == b^l)$, neutral if $1(d_i^l \neq b^l)$ and $1(y \in d_i^l)$, and finally, negative if $1(d_i^l \neq b^l)$ and $1(y \notin d_i^l)$. The overall score for $f_i$ is a weighted sum $S_i$ of all the scores (negative features getting a negative score) along the path. Since we know the locations of the features, we can transfer the scores to actual locations on the images (more than one location may map to the same feature in the vocabulary). When a simple threshold is applied, we get the map seen in the final image. The window most likely to contain the object can then be obtained by optimization of the scores on the map using branch and bound techniques.

## 7. CONCLUSION

We have introduced a novel, biologically derived algorithm that carries out similarity-based hierarchical clustering combined with simple matching, thus determining when nodes in the tree are to be iteratively deepened. The clustering mechanism is a reduced subset of published hypotheses of thalamocortical function; the match/mismatch operation is a reduced subset of proposed basal ganglia operation; both are described in Granger (2006). The resulting algorithm performs a range of tasks, including identifying natural underlying structure among object in the dataset; these abilities of the algorithm confer a range of application capabilities beyond traditional classifiers. In the present paper we described in detail just one circumscribed behavior of the algorithm: its ability to use its combination of unsupervised clustering and reinforcement to carry out the task of supervised classification. The experiments reported here suggest the algorithm's performance is comparable to that of SVMs on this task, yet requires only a fraction of the resources of SVM or KNN methods.

It is worth briefly noting that the intent of the research described here has not been to design novel algorithms, but rather to educe algorithms that may be at play in brain circuitry. The two brain structures referenced here, neocortex and basal ganglia, when studied in isolation, have given rise to hypothesized operations of hierarchical clustering and of reinforcement learning, respectively (e.g., Sutton and Barto, 1998; Rodriguez et al., 2004). These structures are connected in a loop, such that (striatal) reinforcement learning can be hypothesized to selectively interact with (thalamocortical) hierarchies being constructed. We conjecture that the result is a novel composite algorithm (CSL), which can be thought of as iteratively constructing rich representations of sampled data.

Though the algorithm was conceived and derived from analysis of cortico-striatal circuitry, the next aim was to responsibly analyze its efficacy and costs and compare it appropriately against other competing algorithms in various domains. Thus we intentionally produced very general algorithmic statements of the derived cortico-striatal operations, precisely so that (1) we can retain functional equivalency with the referenced prior literature (Schultz et al., 1997; Suri and Schultz, 2001; Schultz, 2002; Rodriguez et al., 2004; Daw and Doya, 2006); and (2) the derived algorithm can be responsibly compared directly against other algorithms. The algorithm can be applied to a number of tasks; for purposes of the present paper we have focused on supervised classification (though we also briefly demonstrated the utility of the method for different tasks, including identification of structure in data, and localization of objects in an image).

It is not yet known what tasks or algorithms are actually being carried out by brain structures. Brain circuits may represent compromises among multiple functions, and thus may not outperform engineering approaches to particular specialized tasks (such as classification). In the present instance, individual components are hypothesized to perform distinct algorithms, hierarchical clustering and reinforcement learning, and the interactions between those components perform still another composite algorithm, the CSL method presented here. (And, as mentioned, the studied operations are very-reduced subsets of the larger hypothesized operations of these thalamocortical and basal ganglia systems; it is hoped that ongoing study will yield further algorithms arising from richer interactions of these cortical and striatal structures, beyond the reduced simplifications studied in the present paper.) As might be expected of a method that has been selectively developed in biological systems over evolutionary time, these component operations may represent compromises among differential selectional pressures for a range of competing tasks, carried out by combined efforts of multiple distinct engines of the brain. This represents an instance in which models of a biological system lead to derivation of tractable algorithms for real-world tasks. Since the biologically derived method studied here substantially outperforms extant engineering methods in terms of efficacy per time or space cost, we forward the conjecture that brain circuitry may continue to provide a valuable resource from which to mine novel algorithms for challenging computational tasks.

## ACKNOWLEDGMENTS

This work was supported in part by grants from the Office of Naval Research and the Defense Advanced Research Projects Agency.

**Conflict of Interest Statement:** The authors declare that the research was conducted in the absence of any commercial or financial relationships that could be construed as a potential conflict of interest.

*Received: 27 May 2011; accepted: 28 October 2011; published online: 10 January 2012.*

*Citation: Chandrashekar A and Granger R (2012) Derivation of a novel efficient supervised learning algorithm from cortical-subcortical loops. Front. Comput. Neurosci.* ***5****:50. doi: 10.3389/fncom.2011.00050*

*This article was submitted to Frontiers in Iterative computations of cortico-striatal loops, a specialty of Frontiers in Computational Neuroscience.*